\documentclass[aps,prl,reprint,superscriptaddress,secnumarabic,amssymb]{revtex4-1}
\usepackage{graphicx}
\usepackage{dcolumn}
\usepackage{bm}
\usepackage{color}
\usepackage[breaklinks=true,colorlinks,citecolor=blue,linkcolor=blue,urlcolor=blue]{hyperref}

\newcommand{\Tr}{\text{Tr}}

\newcommand{\fref}[1]{Fig.\hspace{0.025in}\ref{#1}}

\newcommand{\eref}[1]{Eq.\hspace{0.025in}(\ref{#1})}

\usepackage{graphicx}
\usepackage[normalem]{ulem}
\usepackage{color}
\usepackage{amsmath}
\usepackage{enumitem}

\begin{document}
\title{Weak Values from Displacement Currents in Multiterminal Electron Devices}

\author{D. Marian}
\affiliation{Dipartimento di Fisica dell'Universit\`a di Genova and INFN sezione di Genova, Via Dodecaneso 33, 16146 Genova, Italy}
\affiliation{Departament d'Enginyeria Electr\`onica, Universitat Aut\`onoma de Barcelona, 08193-Bellaterra (Barcelona), Spain}
\author{N. Zangh\`i}
\affiliation{Dipartimento di Fisica dell'Universit\`a di Genova and INFN sezione di Genova, Via Dodecaneso 33, 16146 Genova, Italy}
\author{X. Oriols}
\email{xavier.oriols@uab.cat}
\affiliation{Departament d'Enginyeria Electr\`onica, Universitat Aut\`onoma de Barcelona, 08193-Bellaterra (Barcelona), Spain}

\begin{abstract}
Weak values allow the measurement of observables associated with noncommuting operators. Up to now, position-momentum weak values have been mainly developed for (relativistic) photons. In this Letter, a proposal for the measurement of such weak values in typical electronic devices is presented. Inspired by the Ramo-Shockley-Pellegrini theorem that provides a relation between current and electron velocity,  it is shown that the displacement current measured in multiterminal configurations can provide either a \emph{weak} measurement of the momentum or \emph{strong} measurement of position. This proposal opens new opportunities for fundamental and applied physics with state-of-the-art electronic technology. As an example, a setup for the measurement of the Bohmian velocity of (nonrelativistic) electrons is presented and tested with numerical experiments.   
\end{abstract}


\maketitle


\emph{Introduction.}---Nowadays, there is a rapidly growing interest in weak measurements and weak values \cite{AAV_1988,Nori_2012,Braveman_2013,Dressel_2014}, both from fundamental and applied points of view. Since weak values (a weak measurement postselected by a strong measurement) provide information on incompatible observables associated with noncommuting operators, relevant topics of quantum mechanics, such as the tunneling times \cite{Vona},  Hardy's paradox \cite{Lundeen_2009,Yokota_2009}, Leggett-Garg inequalities \cite{Palacios_2010,Dressel_2011}, and quantum amplification \cite{Hosten_2008,Dixon_2009,Starling_2010}, are being revisited. Especially attractive is the simultaneous measurement of position and momentum: a set of weak measurements of position postselected by a strong measurement of momentum is proportional to the wave function of the system \cite{Lundeen_2011,Lundeen_2012}, while a weak measurement of momentum postselected by a strong measurement of position gives the local velocity of a quantum particle \cite{Wiseman_2007,Traversa_2013}.

Most experimental techniques for weak values are developed for photons, whose technology is not easily transferable to industry based on electronics. The few proposals dealing with weak measurements in solid-state systems \cite{Romito_2011,Williams_2008,Romito_2008,Gefen_2008,Belzig_2010,Clerk_2012} use particle current measurement (i.e., electron charge detection). Instead, we propose measuring displacement current  (i.e. time dependent variations of the electric field) to get information on the position and momentum of a quantum state. Similar to Landauer's proposal \cite{Landauer_1957} which demonstrates that the measured dc current provides information of the quantum transmission coefficient, here, we show that the \emph{weak} measurement of the ac current flowing through a (properly prepared multiterminal) electron device provides information on the whole quantum state. This new proposal opens original routes to study, both, fundamental physics and quantum engineering. 

As an example of the potentialities of our proposal, inspired by the old classical works of Shockley and Ramo \cite{Shockley_1938,Ramo_1939}, we discuss the measurement of the local (Bohmian) velocity (i.e. the current density divided by the modulus of the wave function) for an electron. Such velocity is  obtained from a weak value constructed from two measurements of the displacement current on two different metallic surfaces belonging to a multiterminal device. The electric field generated by a moving electron (which contains information on the electron dynamics) is detected in a large metallic surface even when the electron is far from that surface. Such metallic surface (i.e. the electrons inside) does only weakly perturb the quantum state of the electron. A strong measurement of position can be envisioned by using a small surface that only detects the electric field when the electron is very close. Next, before describing the simpler strong measurement for postselection, we explain the weak measurement.

\begin{figure}[h!!!]
\centering
\includegraphics[width=0.87\columnwidth]{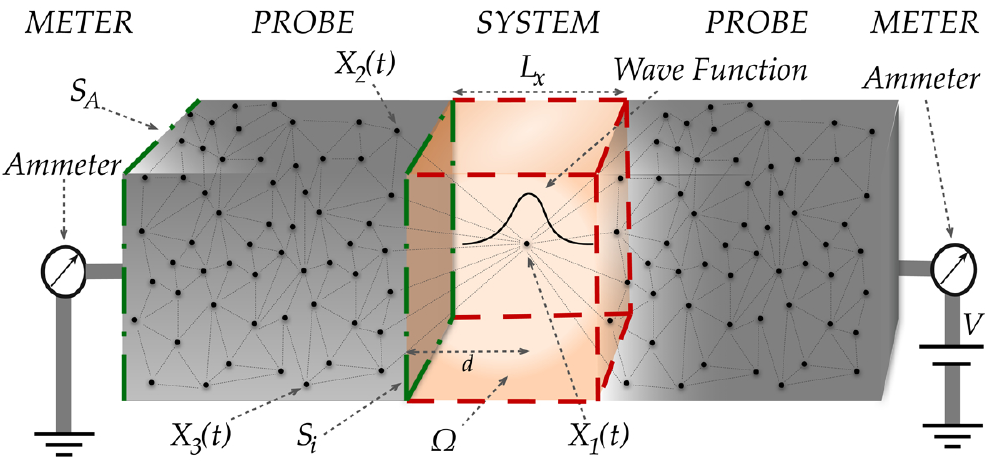}
\caption{Schematic representation of the whole setup divided into three regions. The \emph{system} is enclosed in the volume $\Omega$ (red dashed lines) with an electron inside. The left and right \emph{probes} (cables) with $N_P$ electrons ensuring that the total current on the left surface $S_i$ is equal to that on the left $S_A$ (green dashed-dotted lines). Finally, the left and right \emph{meters} (ammeters) indicate the measured value of the total current (the right probe and meter will be substituted in \fref{fig3}(b) by a multiterminal structure).}
\label{fig1-bis}
\end{figure}

\emph{Weak measurement of the total current.}---The measurement of the electrical current can be understood as a two step process. The first step is an electromagnetic propagation of the total current along the cable (that connects the quantum system and the ammeter in \fref{fig1-bis}). The total current through $S_i$  is equal to the current through $S_A$ far from the active region. This equivalence (due to the divergencelessness of the total current) is exact for the sum of the particle plus the displacement currents, but not for particle current alone. The second step is done by the ammeter that transforms the total current $S_A$ into a pointer value.

In this Letter we analyze this two-step many-body quantum measurement with the corresponding quantum errors and backaction \cite{footnote2}. We define the density matrices of the system and the probe, at the initial time $t_0$, as $\hat{\rho}_{sys}(t_0)$ and $\hat{\rho}_{pro}(t_0)$, respectively.  The time evolution of the density matrix of the entangled system between the initial time $t_0$ and the final time $t_m$ is:
\begin{eqnarray}
 \hat{\rho}_{tot}(t_m)= \hat{U}(t_0,t_m)\left (\hat{\rho}(t_0)_{sys} \otimes \hat{\rho}(t_0)_{pro}\right) \hat{U}(t_0,t_m)^{\dagger},
\label{interact}
\end{eqnarray}
where the unitary operator $\hat{U}(t_0,t_m)$, which contains the free part of the system and of the probe as well as the electromagnetic interaction between them, is the responsible of the first step, i.e., translating the total current from $S_i$ towards $S_A$ (see \fref{fig1-bis}).
  
The second step of the measurement is done at time $t_m$ by a projective measurement of the probe performed by the ammeter. Such measurement provides the value $\tilde{I}$ on the pointer \cite{footnote1}. 
Thus, the probability of the specific result $\tilde{I}$ associated with the eigenstate $|\tilde{I}\rangle$ in the ammeter at time $t_m$ is:
\begin{eqnarray}
\mathcal{P}(\tilde{I},t_m) = \Tr_{pro} \left( | \tilde{I} \rangle \langle \tilde{I} | \hat{\rho}'_{pro}(t_m)\right),
\label{partialpro}
\end{eqnarray}
where $\hat{\rho}'_{pro}(t_m)=\Tr_{sys}(\hat{\rho}_{tot}(t_m))$ is the reduced density matrix of the probe, with  $\Tr_{sys}$ the partial trace operation over the system coordinates. Hereafter, we will consider only one particle in the system, while considering an arbitrary large number $N_P$ of particles in the probe.  

The standard way of describing the probability $\mathcal{P}(\tilde{I},t_m)$ is not by referring to the whole probe\rq{}s coordinates in \eref{partialpro}, but only to the quantum system\rq{}s coordinates: 
\begin{eqnarray}
\mathcal{P}(\tilde{I},t_m) =  \Tr_{sys} \left( \hat{I}(\tilde{I}) \hat{\rho}'_{sys}(t_m)\right),
\label{partialpro2}
\end{eqnarray}
where $\hat{\rho}'_{sys}(t_m)=\Tr_{pro}(\hat{\rho}_{tot}(t_m))$ is the system reduced density matrix, with $\Tr_{pro}$ the partial trace operation over the probe\rq{}s coordinate. In \eref{partialpro2}, we define $\hat{I}(\tilde{I})$ as a general positive operator-valued measure (POVM): 
\begin{eqnarray}
\hat{I}(\tilde{I}) = \int | I \rangle g(\tilde{I},I) \langle I | dI,
\label{operator}  
\end{eqnarray}
with $g(\tilde{I},I)$ a real positive number. If we introduce the quantum system density matrix $\hat{\rho}'_{sys}(t_m)=\sum_J p_J(t_m) | \psi_J \rangle \langle \psi_J|$  with $p_J(t_m)$ a normalized probability and $|\psi_J\rangle=\int a^J_I(t_m) |I\rangle dI$ into \eref{partialpro2}, together with \eref{operator}, the term $g(\tilde{I},I)$  can be computed from the relation: 
\begin{eqnarray}
\mathcal{P}(\tilde{I},t_m) = \sum_J p_J(t) \int g(\tilde{I},I) |a^J_I(t_m)|^2 dI.
\label{relation}  
\end{eqnarray}
Next, we discuss  in what conditions the measurement of the total current provides information on the momentum of the quantum system.


\emph{Displacement current and momentum measurement}.---The (quantum ensemble) value of the total current $\langle I \rangle$ can be computed \cite{Pellegrini_1986,Albareda_2012} straightforwardly as the sum of particle plus displacement current on the surface $S_i$:
\begin{eqnarray}
&&\langle I(t) \rangle  = \int_{S_i}  \langle{ \mathbf{J}_{c}(\mathbf{r},t) }\rangle \cdot d\mathbf{s} + \int_{S_i}  \epsilon \frac{d\langle{\mathbf{E}(\mathbf{r},t)}\rangle}{dt} \cdot d\mathbf{s},
\label{tc_gen}
\end{eqnarray}
where $\epsilon$ is the dielectric constant, $ \langle{J}_{c}(\mathbf{r},t) \rangle$ the quantum ensemble value of the quantum particle current density and $\langle{\mathbf{E}(\mathbf{r},t)}\rangle$ the quantum ensemble value of the electric field on the surface. Identically, from the  Ramo-Schockley-Pellegrini theorem with a quasistatic approximation that neglects the vector potential (magnetic) contribution \cite{footnote7,Pellegrini_1986,Albareda_2012,Electric_Field}, such a mean value in \eref{tc_gen} can be written as
\begin{eqnarray}
\!\!\!\!\!\langle I(t) \rangle  \!\!=\!\!  - \! \!\int_{\Omega}\! \!\mathbf{F}(\mathbf{r}) \!\cdot \!\!\langle{ \mathbf{J}_{c}(\mathbf{r},t) }\rangle\! \cdot \!dv \!+ \!\!\!\int_{S}\!\! \epsilon \!\cdot \!\mathbf{F}(\mathbf{r}) \!\cdot \!\frac{d \langle{ V(\mathbf{r},t) }\rangle}{dt}  \!\cdot\! d\mathbf{s},
\label{tc_rst}
\end{eqnarray}
where $S$ is the close surface of the volume $\Omega$ (see~\fref{fig1-bis}) that contains $S_i$ and $\langle{ V(\mathbf{r},t) }\rangle$ is the ensemble value of the electrostatic potential. In the particular situation where electron transport takes place between two metallic surfaces included into $\Omega$, no variation of the potential appears on these surfaces, i.e., $d\langle{V(\mathbf{r},t)}\rangle/dt = 0$. If the square of the distance between the metallic surfaces is smaller than the area of the surfaces, $L_x^2 \ll S_i$,  we get $F(\mathbf{r})_x=1/L_x$ \cite{Pellegrini_1986,Abdelilah_2012}. Then, \eref{tc_rst} is rewritten as
\begin{equation}
\langle I(t) \rangle = \frac{q}{m L_x}\langle p(t) \rangle_{\Omega},
\label{RSt_Metal}
\end{equation}
where $\langle p(t) \rangle_{\Omega}$ is the mean value of the momentum in the volume $\Omega$ of the device. We have assumed that the support of the density matrix of the quantum system is inside the volume $\Omega$. See Ref. \cite{footnote2} for an alternative demonstration of \eref{RSt_Metal}. The experimental evaluation of this mean momentum requires the knowledge of the probabilities in \eref{relation}.


\emph{Numerical evaluation of $\mathcal{P}(\tilde{I},t)$.}---In order to compute the probability $\mathcal{P}(\tilde{I},t)$, one needs to simulate the time evolution in \eref{interact} and use \eref{partialpro}. Obviously, the exact solution of this problem is not accessible, but fortunately we can provide some reasonable simplifications to handle this problem. The many-particle Coulomb interaction among electrons is enough to consider the transmission of the total current from $S_i$ towards $S_A$ \cite{footnote7,Albareda_2012,footnote3}. 

To numerically treat the problem we use the method reported in Ref.~\cite{Oriols} based on the definition of \emph{conditional wave functions} $\psi^j_i(x_i,t)=\Psi(\mathbf{X}^j_1(t),...,\mathbf{X}^j_{i-1}(t),x_i,\mathbf{X}^j_{i+1}(t),...,t) $ for each $i$-particle with $\Psi$ the many particle wave function \cite{Durr_2004,Durr_1992,Durr_2007,Oriols}. For simplicity, the quantum system is treated as a 1D system. The capital letter $\mathbf{X}^j_i(t)$ denotes the actual (Bohmian) particle 3D position. The subindex $i=1,...,N_P+1$ refers to the particle in the quantum system plus the $N_P$ particles in the probe. The superindex $j=1,...,M$ denotes one (of the infinite) particular selections of the initial positions. The time evolution of $\psi^j_1(x_1,t)$ is obtained by solving numerically the following single-particle (conditional) Schr\"odinger-type equation:
\begin{equation}
\label{eq-conditional}
i\hbar \frac{\partial \psi^j_1(x_1,t)}{\partial t} = \left[ H_0 + V \right] \psi^j_1(x_1,t) 
\end{equation}
where $V = V(x_1,\mathbf{X}^j_2(t),...,\mathbf{X}^j_{N_P+1}(t))$ is the conditional Coulomb potential felt by the system and $H_0$ is its free Hamiltonian. The trajectory $X^j_1(t)$ is obtained from the so-called \emph{guidance equation} $v^j_1(t)=dX^j_1(t)/dt=(\hbar/m) \text{Im}\big((\partial \psi^j_1/\partial x_1) (1/\psi^j_1)\big)$. This numerical method \cite{Oriols} for our problem allows the following: (i) A manageable treatment of the many-particle interaction of \eref{interact} for a small interaction \cite{PRB} and (ii) a simple treatment of  \eref{partialpro} by using, for each experiment, one \emph{channelized} conditional wave function  $\Psi(x_1,\mathbf{X}^j_{2}(t),...,\mathbf{X}^j_{N_P+1}(t),t)$ \cite{Durr_1992}.
The quantum probabilities computed from Bohmian trajectories are, by construction, identical to the ones obtained from standard quantum tools. Once an (infinite) $j=1,...,M$ ensemble of initial positions are considered, we compute the probability $\mathcal{P}(\tilde{I},t)$ in the range $[\tilde{I} ,\tilde{I}+{d\tilde{I}}]$ as $\mathcal{P}(\tilde{I},t) d\tilde{I} = \sum_{j=1}^{M} \Theta(\tilde{I}^j(t)-\tilde{I}) \Theta(\tilde{I}+{d\tilde{I}}-\tilde{I}^j(t))$, with $\Theta(x)$ the Heaviside function. The measured total current $\tilde {I}^j(t)$ is computed as:
\begin{eqnarray}
\label{eq-1}
\tilde {I}^{j}(t)  =  \int_{S_i}  \epsilon \frac{d\mathbf{E}^j}{dt} \cdot d\mathbf{s} \!= \! \sum_{k=1}^{1+N_P} \epsilon \mathbf{\nabla} \Phi(\mathbf{X}^j_k(t)) \cdot \mathbf{v}^j_k(t),
\end{eqnarray}
where $\mathbf{E}^j$ is the electric field generated on the surface $S_i$ by the $1+N_P$ particles $\mathbf{X}^j_k$. We include the previous \emph{conditional} wave function algorithm in the right-hand side of \eref{eq-1} where $\Phi$ is the electric flux due to the $i$th electron and $\mathbf{v}^j_i(t)$ is the local (Bohmian) velocity. We shall use \eref{eq-1} as a numerical tool for calculating the subsequent results (see details in Ref. \cite{footnote2}).

\begin{figure}[h!!!]
\centering
\includegraphics[width=0.95\columnwidth]{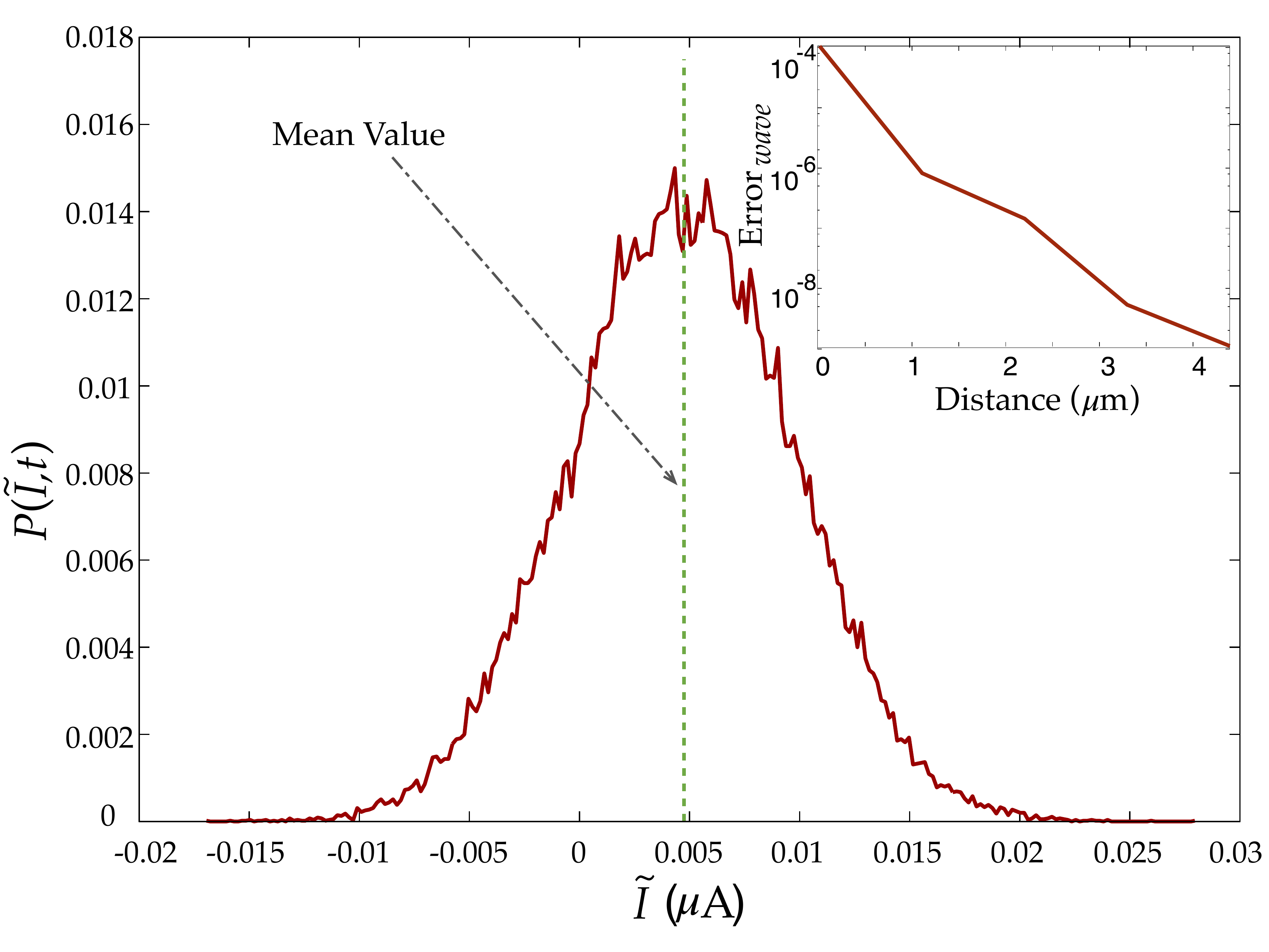}
\caption{Probability distribution (red solid line) of the measured total current from \eref{eq-1}  from $M = 55000$ numerical experiments. Mean value of the distribution (green dashed line) obtained from the same simulations without the measurement backaction. Inset: Error on the wave function computed as Error$_{wave} = \int |\psi(x_1,t_m)-\psi^{mean}(x_1,t_m)|^2 dx_1$ at the final time $t_m$ for different values of the distance $d$ in \fref{fig1-bis}. The wave function $\psi(x_1,t_m)$ is computed from \eref{eq-conditional} with the many-particle potential, while $\psi^{mean}(x_1,t_m)$ from \eref{eq-conditional} with only an external mean-field potential.}
\label{fig2}
\end{figure}

In \fref{fig2} we show the numerical results of $\mathcal{P}(\tilde{I},t_m)$ when $M=55000$ for a \emph{large} surface $S_A$, i.e., for $S_A\gg L_x^2$. For simplicity,  the electron $1$ in the quantum system is moved only in the 1D transport direction $x$ through \eref{eq-conditional} interacting, through Coulomb potentials, with the other electrons in the left and right probes. The $N_P$ electrons are simulated semiclassically in a 3D space taking into account the many-body Coulomb interaction among them (and with the particle in the system) plus the interaction with a bath of phonons at room temperature. Although we obtain a large dispersion of values of $\tilde{I}$, its mean value exactly coincides with the mean value $\langle I  \rangle$ obtained without including the $N_P$ electrons in the metals. In addition, we observe that the system is only slightly modified by the interaction with the electrons in the metal when the distance is reduced (see inset in \fref{fig2}). 

We use \eref{relation} to find $g(\tilde{I},I)$. The support of the function  $\mathcal{P}(\tilde{I},t)$ is much larger than the current (momentum) distribution of the wave function, $|a^J_I(t)|^2$ for any $J$. This information can be obtained comparing the distribution $\mathcal{P}(\tilde{I},t)$ in \fref{fig2} and the $|a^J_I(t)|^2$, which are known directly from the wave function $|\psi_J\rangle$. Therefore, we can approximate $|a^J_I(t)|^2 \approx \delta(\langle I \rangle-I)$ in \eref{relation} to get $g(\tilde{I},\langle I \rangle) \approx \mathcal{P}(\tilde{I},t_m)$. We have used $\sum_J p_J(t)=1$. From these last results, along with \fref{fig2} and \eref{RSt_Metal}, we obtain a specific expression for the 1D version of the operator in \eref{operator}:
\begin{equation}
\hat{I}_{w}(p_w) \approx C_w \int dp\; e^{-\frac{(p-p_w)^2}{2\sigma_w^2}} |p\rangle \langle p|,
\label{weak-def-I-text} 
\end{equation}
where $\sigma_w$ is the width of the Gaussian distribution in \fref{fig2} (related to the specific plasma oscillations of the metal probes) and $C_w$ is a suitable normalization constant. The subindex $w$ reflects the particular geometry of the device that leads to a weak measurement. 

\begin{figure}[h!!!]
\centering
\includegraphics[width=0.90\columnwidth]{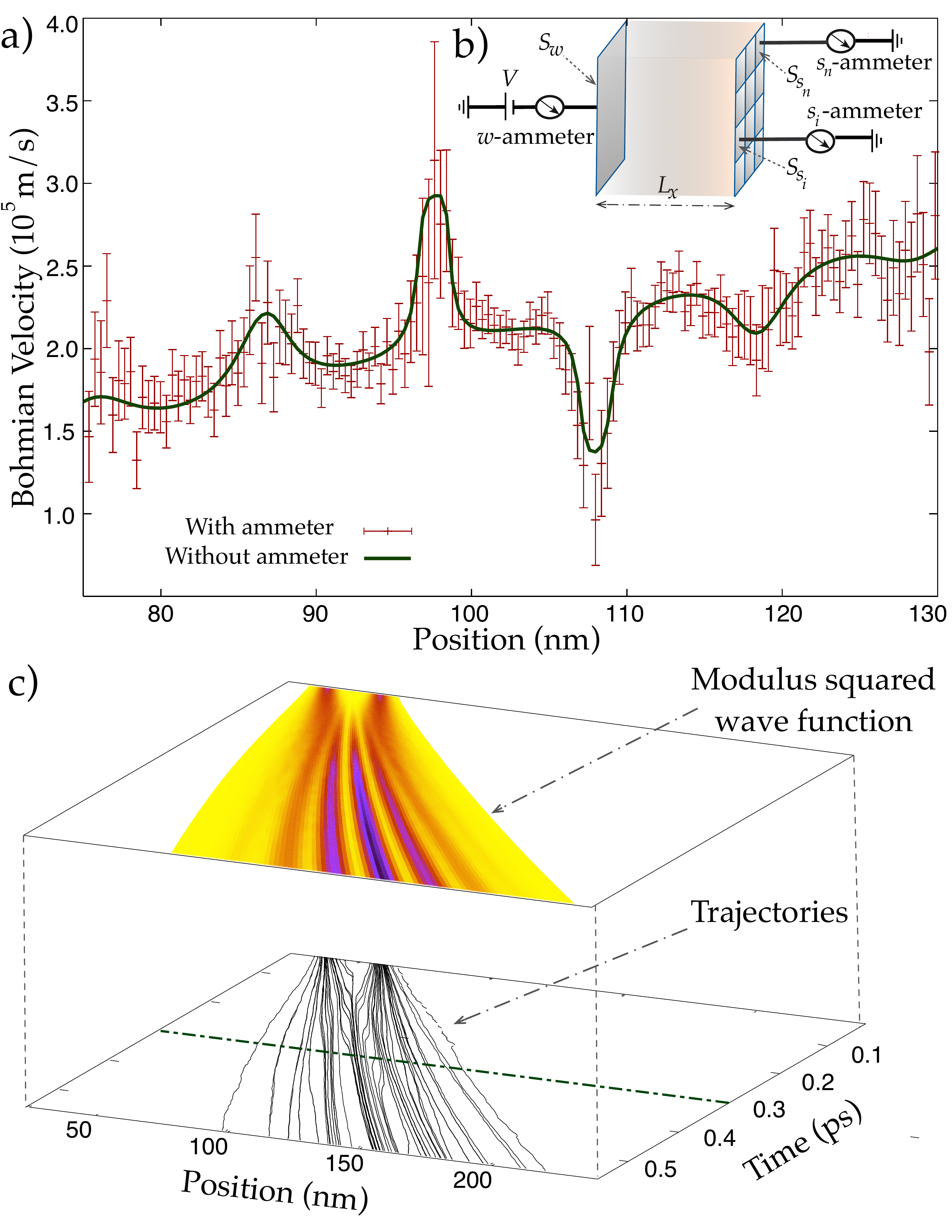}
\caption{(a) Bohmian velocity (red points with errors) from \eref{eq-bv-final-text} as a function of position $x_s$ at time $t_m = 0.3 \; ps$, obtained from an ensemble of numerical experiments with the backaction of the ammeter. Its errors are calculated from the standard deviation of the set of velocities at time $t_m$ and position $x_s$ divided by the square root of the number of trajectories passing through position $x_s$ at time $t_m$. The Bohmian velocity (green solid line) in position $x_s$ obtained from the same set of simulations without the backaction. (b) Schematic representation of the multiterminal device described in the text where $S_w$ and different $S_{s_i}$ are indicated. (c) Wave function (upper plot) obtained from simulations without the backaction of the measurement and Bohmian trajectories (lower plot) reconstructed from the $M$ experiments with backaction, at different times. The green dashed line represents the time chosen for the plot in a). }
\label{fig3}
\end{figure}

\emph{Measurement of Bohmian velocities.}---Hereafter, to show the possibilities of these weak measurements of the displacement current, we provide a proposal for measuring the Bohmian velocity of a quantum particle. We consider a multiterminal device where there are two metallic surfaces working as \emph{sensing electrodes}. See \fref{fig3}(b). The length of the device is $L_x = 280\;nm$ and the surface $S_{w} \approx 10^{-11}\;m^2$, satisfying the relation $S_w \gg L_x^2$. The total current can be obtained from the operator in \eref{weak-def-I-text}. The other \emph{electrode} is divided into $n$ surfaces, each one electrically insulated from the others, connected to its $s_i-$ammeter and satisfying the opposite relation $S_{s_i} \ll L_x^2$. In these small surfaces the total current in \eref{tc_rst} can be computed from $F(\mathbf{r})_x  = -\alpha e^{-\alpha (|\mathbf{r}-\mathbf{r_s}|)}$  with $\alpha = \sqrt{2/S_{s_i}}$ and $\mathbf{r_s}=\{x_s,y_s,z_s\}$ the central position of  $S_{s_i}$ \cite{Abdelilah_2012}. Here,  \eref{tc_rst} provides a measurement of the total current that becomes different from zero when the particle is going to cross one of the small surfaces of \fref{fig3}(b). In this small surface, the 1D version of the total current operator in \eref{operator} is:
\begin{equation}
\hat{I}_{s}(x_s) \approx C_s \int dx e^{-\frac{(x-x_s)^2}{2\sigma_s^2}} |x\rangle \langle x|.
\label{strong-def-I-text} 
\end{equation}
where $\sigma^2_s$ is related to $S_{s_i}$ in \fref{fig3}(b). See Ref. \cite{footnote2} for an alternative derivation of \eref{strong-def-I-text}.

The exact procedure for measuring the Bohmian velocity is the following: The total density matrix evolves from $t_0$ till $t_m$ following \eref{interact}. At time $t_m$ the $w$-ammeter measures weakly the momentum of the particle through \eref{partialpro2}. Then, the quantum system evolves until a peak of current is measured (or not) in the $s_i$-ammeter. The values measured with the $w$-ammeter will be postselected by the measurement of the positions performed by the $s_i$-ammeter. Thus, we compute
\begin{eqnarray}
E[p_w|x_s] = \frac{\int dp_w p_w \mathcal{P}(p_w \cap x_s)}{\mathcal{P}(x_s)},
\label{prima-text}
\end{eqnarray}
which under the condition $\sigma_w \gg (\hbar/\sigma_s)$ becomes
\begin{eqnarray}
\frac{E[p_w|x_s]}{m} = \frac{J(x_s,t_m)}{|\psi(x_s,t_m)|^2} \equiv v(x_s,t_m)
\label{eq-bv-final-text}
\end{eqnarray}
where $m$ is the particle mass and $v(x_s,t_m)$ is exactly the Bohmian velocity. See Ref. \cite{footnote2} for an explicit development from \eref{prima-text} till \eref{eq-bv-final-text}.

Hereafter, to provide a realistic estimation on how many experiments $M$ are needed to capture typical quantum interference phenomena in electronic devices and to understand the backaction in the system, we provide a direct numerical simulation of the whole weak value procedure by simulating \eref{interact} from $t_0$ till $t_m$ and a posterior measurement through \eref{partialpro2} using the \emph{conditional wave function} technique mentioned in \eref{eq-conditional}, without using the operators defined in \eref{weak-def-I-text} and \eref{strong-def-I-text}.
In \fref{fig3}(a) it is reported the Bohmian velocity $v(x_s,t)$ obtained from an ensemble of $M=55000$ (identically prepared) two-time measurement experiments calculated from \eref{eq-bv-final-text}. The different errors are due to the different number of particles that effectively arrive at each position. In order to see interference effects we consider as initial wave function $\psi(x_1,0)$ a superposition of two Gaussian wave packets, in the motion direction $x$, whose central positions are separated of $50\;nm$  at the initial time (see \fref{fig3}c). Each Gaussian wave packet has the same dispersion of $3\;nm$ and energy of $0.0905\;eV$. The velocity field exhibits the typical interference pattern. 

In \fref{fig3}(c) we report the comparison of the trajectories obtained from the described procedure and the wave function of the single particle problem. As expected, the trajectories are more dense near the maximum and less dense near the minimum of the interference pattern.  


\emph{Conclusions.}---In conclusion, we have presented an experimental proposal for measuring weak values of position and momentum in solid state devices. In particular, we show how the Bohmian velocity of a (massive) electron can be obtained from weak values of the displacement currents in a multiterminal electron device. We emphasize that the Wiseman\rq{}s protocol \cite{Wiseman_2007} for (weak) measuring the Bohmian velocity was developed for massive nonrelativistic particles (not for massless relativistic ones). So, our Bohmian velocity measurement for electrons in solid-state structures exactly fulfills the nonrelativistic scenario contemplated in Ref. \cite{Wiseman_2007}, while the local velocity of an ensemble of relativistic photons measured in the pioneering work of Kocsis \emph{et al.} \cite{Kocsis_2011} does not. The feasibility of our proposal has been tested numerically in the iconic double-slit experiment at room temperature. In real experiments, for samples at low temperatures with few microns coherence length \cite{coherence}, the protocol presented here can be implemented with frequencies below $50 \;GHz$. Our protocol implies the use of the (now available) single electron sources \cite{Boc,New}, while standard thermal injection in solid-state devices would provide the (ensemble) velocity of the mixed state \cite{Angel}. This work opens the path for answering intricate fundamental questions and developing new quantum engineering applications using the successful electronic/semiconductor industry.

\begin{acknowledgments}
The authors would like to thank Tom\'as Gonz\'alez, Javier Mateos, Philippe Dollfus and Massimo Macucci for fruitful discussions and suggestions. This work has been partially supported by the Fondo Europeo de Desarrollo Regional (FEDER) and Ministerio de Econom'a y Competitividad through the Spanish Projects No. TEC2012-31330 and No. TEC2015-67462-C2-1-R, the Generalitat de Catalunya (2014 SGR-384), and by the European Union Seventh Framework Program under the Grant Agreement No. 604391 of the Flagship initiative ``Graphene-Based Revolutions in ICT and Beyond". N. Z. is supported in part by INFN.
\end{acknowledgments}

\newpage

\section{A - Derivation of the Bohmian Velocity from Kraus operators}
\label{app1}

Alternatively to the numerical experiment developed in the letter (see Fig. 3 in the letter), hereafter we provide a detailed analytical derivation of Eq. (14) using the Kraus operators $\hat I_w$  and $\hat I_s$ defined in the letter, Eq. (11) and Eq. (12), respectively. We rewrite the final expression here as:

\begin{eqnarray}
\!\!\!\!v(x_s,t_m)\! =\! \frac{E[p_w|x_s]}{m} \!=\! \frac{1}{m} \frac{\int dp_w p_w \mathcal{P}(p_w \cap x_s)}{\mathcal{P}(x_s)}.
\label{eq-bv-final-app}
\end{eqnarray}

First, we calculate the probability ${\mathcal{P}(x_s)}$ in denominator of the last expression as:

\begin{eqnarray}
\mathcal{P}(x_s)  &=& \int  dp_w \mathcal{P}(p_w \cap x_s) \nonumber \\
&=& \int dp_w dp_w \langle \psi | \hat{I}^{\dagger}_{w} \hat{U}^{\dagger}_{t_m} \hat{I}^{\dagger}_{s}  \hat{I}_{s} \hat{U}_{t_m} \hat{I}_{w} |\psi \rangle \nonumber \\
&=&   \iint  dp' dp'' \left[ C_w^2  \int dp_w e^{-\frac{(p'-p_w)^2}{2\sigma_w^2}} e^{-\frac{(p''-p_w)^2}{2\sigma_w^2}}\right] \cdot \nonumber \\ 
&\cdot& \langle \psi | p' \rangle \langle p' | \hat{U}^{\dagger}_{t_m} \hat{I}^{\dagger}_{s}  \hat{I}_{s} \hat{U}_{t_m} | p'' \rangle \langle p'' | \psi \rangle, 
\label{eq-primo}
\end{eqnarray}

where in the last line it has been used the definition in Eq. (11) and Eq. (12). We can focus only in the integral between squared parenthesis in the last line of \eref{eq-primo}

\begin{eqnarray}
&&C_w^2 \int dp_w e^{-\frac{(p'-p_w)^2}{2\sigma_w^2}} e^{-\frac{(p''-p_w)^2}{2\sigma_w^2}} = \nonumber \\
&=& C_w^2 \int dp_w e^{-\frac{p_w^2}{\sigma_w^2}+p_w\left( \frac{p'}{\sigma_w^2}+\frac{p''}{\sigma_w^2}\right)-\frac{p'^2}{2\sigma_w^2}-\frac{p''^2}{2\sigma_w^2}}= \nonumber \\
&=& e^{-\frac{(p'-p'')^2}{4\sigma_w^2}}
\end{eqnarray}

where it has been used $C_w=(\sqrt{\pi}\sigma_w)^{-1/2}$. Thus \eref{eq-primo} becomes:

\begin{eqnarray}
&&\mathcal{P}(x_s) = \nonumber \\
&=&  \iint dp' dp'' e^{-\frac{(p'-p'')^2}{4\sigma_w^2}}\langle \psi | p' \rangle \langle p' | \hat{U}^{\dagger}_{t_m} \hat{I}^{\dagger}_{s}  \hat{I}_{s} \hat{U}_{t_m} | p'' \rangle \langle p'' | \psi \rangle \nonumber \\
&=&  \iint dp' dp'' \langle \psi | p' \rangle e^{-\frac{(p'-p'')^2}{4\sigma_w^2}} \langle p'' | \psi \rangle \cdot \nonumber \\
&\cdot& \left[ C_s^2 \int dx e^{-\frac{(x-x_s)^2}{\sigma_s^2}} \langle p' |  \hat{U}^{\dagger}_{t_m} | x \rangle \langle x | \hat{U}_{t_m} | p'' \rangle \right]. 
\label{eq-secondo}
\end{eqnarray} 

Recalling that:

\begin{itemize}
\item (i) $\hat{U}_{t_m} = \int dp |p\rangle \langle p| e^{-i\frac{p^2t_m}{2m\hbar}}$,
\item (ii) $\langle x | p \rangle = \frac{1}{\sqrt{2\pi\hbar}}e^{i\frac{px}{\hbar}}$,
\item (iii) $\langle x | \hat{U}_{t_m} | p'' \rangle = \langle x | \int dp  |p\rangle \langle p| e^{-i\frac{p^2t_m}{2m\hbar}} | p'' \rangle = \frac{1}{\sqrt{2\pi\hbar}} e^{-i\frac{p''^2t_m}{2m\hbar}+i\frac{p''x}{\hbar}}$,
\end{itemize}

it is possible to work out the integral between squared parenthesis in \eref{eq-secondo}, 

\begin{eqnarray}
&&C_s^2 \int dx e^{-\frac{x^2}{\sigma_s^2}+x\left( \frac{2x_s}{\sigma_s^2} + i\frac{p''}{\hbar}-i\frac{p'}{\hbar} \right)-\frac{x_s^2}{\sigma_s^2} -i\frac{p''^2t_m}{2m\hbar}+i\frac{p'^2t_m}{2m\hbar}} = \nonumber \\
&=& e^{-\frac{\sigma_s^2}{4\hbar^2}(p'-p'')^2}e^{-\frac{i}{\hbar}x_sp'+i\frac{p'^2t_m}{2m\hbar}}e^{\frac{i}{\hbar}x_sp''-i\frac{p''^2t_m}{2m\hbar}} = \nonumber \\
&=& e^{-\frac{\sigma_s^2}{4\hbar^2}(p'-p'')^2} \langle p' |  \hat{U}^{\dagger}_{t_m} | x_s \rangle \langle x_s | \hat{U}_{t_m} | p'' \rangle.
\end{eqnarray}

Therefore \eref{eq-secondo} becomes:

\begin{eqnarray}
\mathcal{P}(x_s) &=&  \iint  dp' dp'' e^{-\frac{(p'-p'')^2}{4\sigma_w^2}} e^{-\frac{\sigma_s^2}{4\hbar^2}(p'-p'')^2} \cdot \nonumber \\
&\cdot& \langle \psi | p' \rangle  \langle p' |  \hat{U}^{\dagger}_{t_m} | x_s \rangle \langle x_s | \hat{U}_{t_m} | p'' \rangle \langle p'' | \psi \rangle,
\label{eq-terzo}
\end{eqnarray}

if the condition 

\begin{eqnarray}
\sigma_w \gg \frac{\hbar}{\sigma_s},
\label{eq-condition}
\end{eqnarray}

is satisfied then \eref{eq-terzo} simply becomes:

\begin{eqnarray}
\mathcal{P}(x_s) = \langle \psi | \hat{U}^{\dagger}_{t_m} \hat{I}^{\dagger}_{s} \hat{I}_s \hat{U}_{t_m} | \psi \rangle = |\psi(x_s,t_m)|^2.
\label{extra1}
\end{eqnarray}

The second step is the calculation of the numerator of \eref{eq-bv-final-app}:

\begin{eqnarray}
&& \int dp_w p_w \mathcal{P}(p_w \cap x_s) = \nonumber \\
&=&  \iint  dp' dp'' \! \left[ \! C_w^2 \int dp_w p_w e^{-\frac{(p'-p_w)^2}{2\sigma_w^2}} e^{-\frac{(p''-p_w)^2}{2\sigma_w^2}}\right] \cdot \nonumber \\
&\cdot& \langle \psi | p' \rangle \langle p' | \hat{U}^{\dagger}_{t_m} \hat{I}^{\dagger}_{s}  \hat{I}_{s} \hat{U}_{t_m} | p'' \rangle \langle p'' | \psi \rangle,
\label{eq-quarto}
\end{eqnarray} 

again it is possible to calculate the integral between squared parenthesis in the last expression,

\begin{eqnarray}
&&C_w^2 \int dp_w p_w e^{-\frac{(p'-p_w)^2}{2\sigma_w^2}} e^{-\frac{(p''-p_w)^2}{2\sigma_w^2}} = \nonumber \\
&=& e^{-\frac{(p'-p'')^2}{4\sigma_w^2}} C_w^2 \int dp_w p_w e^{-\frac{\left[ p_w - \frac{(p'+p'')}{2} \right]}{\sigma_w^2}} = \nonumber \\
&=& \left( \frac{p'+p''}{2} \right)  e^{-\frac{(p'-p'')^2}{4\sigma_w^2}}.
\end{eqnarray}

Then \eref{eq-quarto} becomes:

\begin{eqnarray}
&&\int dp_w p_w \mathcal{P}(p_w \cap x_s) = \nonumber \\
&=&  \iint dp' dp''  \left( \frac{p'+p''}{2} \right) e^{-\frac{(p'-p'')^2}{4\sigma_w^2}} \cdot \nonumber \\
&\cdot& \langle \psi | p' \rangle \langle p' | \hat{U}^{\dagger}_{t_m} \hat{I}^{\dagger}_{s}  \hat{I}_{s} \hat{U}_{t_m} | p'' \rangle \langle p'' | \psi \rangle,
\label{eq-quinto}
\end{eqnarray}

making the same steps done from \eref{eq-secondo} till \eref{eq-terzo}, \eref{eq-quinto} becomes:

\begin{eqnarray}
&& \int dp_w p_w \mathcal{P}(p_w \cap x_s) = \nonumber \\
&=& \iint  dp' dp'' \left( \frac{p'+p''}{2} \right) \!\! e^{-\frac{(p'-p'')^2}{4\sigma_w^2}} \!\! e^{-\frac{\sigma_s^2}{4\hbar^2}(p'-p'')^2} \cdot \nonumber \\
&\cdot& \langle \psi | p' \rangle  \langle p' |  \hat{U}^{\dagger}_{t_m} | x_s \rangle \langle x_s | \hat{U}_{t_m} | p'' \rangle \langle p'' | \psi \rangle. 
\label{eq-sesto}
\end{eqnarray}

If \eref{eq-condition} is satisfied then \eref{eq-sesto} becomes:

\begin{eqnarray}
&&\int dp_w p_w \mathcal{P}(p_w \cap x_s) = \nonumber \\
&=& \frac{1}{2} \left[ \langle \psi | \hat{p} \hat{U}^{\dagger}_{t_m} \hat{I}^{\dagger}_{s} \hat{I}_s \hat{U}_{t_m} | \psi \rangle + \langle \psi | \hat{U}^{\dagger}_{t_m} \hat{I}^{\dagger}_{s} \hat{I}_s \hat{U}_{t_m} \hat{p} | \psi \rangle\right] = \nonumber \\
&=& \text{Re}\left( \langle \psi | \hat{U}^{\dagger}_{t_m} \hat{I}^{\dagger}_{s} \hat{I}_s \hat{U}_{t_m} \hat{p} | \psi \rangle \right)
\label{eq-settimo}
\end{eqnarray}

where in the last expression it has been used the property $\hat{p} = \int p |p \rangle \langle p | dp$. If one writes the momentum operator in position representation easily realizes that  the expression between parenthesis in the last line of \eref{eq-settimo} is

\begin{eqnarray}
&& \langle \psi | \hat{U}^{\dagger}_{t_m} \hat{I}^{\dagger}_{s} \hat{I}_s \hat{U}_{t_m} \hat{p} | \psi \rangle = -i\hbar \psi^*(x_s,t_m) \frac{\partial}{\partial x_s} \psi(x_s,t_m) \nonumber \\ 
&&
\end{eqnarray}

and thus its real part is

\begin{eqnarray}
&&\text{Re}\left( \langle \psi | \hat{U}^{\dagger}_{t_m} \hat{I}^{\dagger}_{s} \hat{I}_s \hat{U}_{t_m} \hat{p} | \psi \rangle \right) = \nonumber \\
&=& \frac{\hbar}{2i} \left( \frac{\partial \psi(x_s,t_m)}{\partial x_s}\psi^*(x_s,t_m) - \frac{\partial \psi^*(x_s,t_m)}{\partial x_s}\psi(x_s,t_m)\right) \nonumber \\
 &=& m J(x_s,t_m). 
 \label{extra2}
\end{eqnarray}

Therefore, using \eref{extra1} and \eref{extra2},  the right hand side of \eref{eq-bv-final-app} becomes:

\begin{eqnarray}
\frac{1}{m} \frac{\int dp_w p_w \mathcal{P}(p_w \cap x_s)}{\mathcal{P}(x_s)} = \frac{J(x_s,t_m)}{|\psi(x_s,t_m)|^2} = v(x_s,t_m),\nonumber
\end{eqnarray}

which is the result in Eq. (14) of the letter that we wanted to demonstrate.

\section{B - Displacement Current and momentum measurements on a large Surface}
\label{app2}

In the text, we use the quantum version of the Ramo-Shockley-Pellegrini theorems in Eq. (8) to demonstrate that the measurement of the displacement current in a large surface is equal to the momentum measurement. Hereafter, we provide an alternative demonstration of Eq. (9) in the text through the use of the \emph{Conditional} wave function algorithm described in the letter. 

According to our discussion in the letter, the capital letters $\{X^j_i(t),Y^j_i(t),Z^j_i(t)\}$ denotes the actual (Bohmian) positions of the particles, where $i$ identifies the $i-th$ particle and $j$ define one (from the infinite) particular selections of the initial positions. The flux of the electric field through a general ideal surface $S_w$ (see Fig. 1), defined as a plane of area  $L_y \cdot L_z$ perpendicular to the $\hat{x}$ direction and placed in $x = x_{w}$, i.e. defined by the points  $\{x_w,0 \leq y' \leq L_y,0 \leq z' \leq L_z\}$, generated by a particle in position $\{X,Y,Z\}$ can be calculated as:

\begin{equation}
\Phi(X,Y,Z)=\int_{S_{w}} \mathbf{E}(X,Y,Z,x_{w},y\rq{},z\rq{}) \cdot d\mathbf{s},
\label{fflux}
\end{equation}
where we have eliminated the superindex $j$, subindex $i$ and time $t$ to simplify the notation. The electric field $\mathbf{E}$ is just computed from the Coulomb force of the electron in the mentioned surface. In the simple case in which the particle is located in $\{X,L_y/2,L_z/2\}$, and it moves only in the $\hat{x}$ direction and $L_y = L_z \equiv L$  ($S_w = L^2$), \eref{fflux} becomes: 

\begin{eqnarray}
\label{newdisp}
&&\Phi(X) = \frac{q}{\pi\epsilon} \!\! \tan^{-1} \!\! \left( \frac{S_w}{4(x_w-X)\sqrt{(x_w-X)^2+\frac{S_w}{2}}}\right). \nonumber \\
&&
\end{eqnarray}

Let us evaluate \eref{newdisp} in the situation in which $S_w \gg (x_w-X)^2$. This means that the maximum distance (squared) between the electron inside the device active region and  the surface is much smaller than the surface itself. In order to work out an approximate form for \eref{newdisp} in this regime it can be considered the following change of variable $\chi = (x_w-X)$. For simplicity, we assume that the electron is located on the left of the surface (i.e. $X < x_w \rightarrow \chi > 0$) then:  

\begin{eqnarray}
\Phi(\chi)  &=&  \frac{q}{\pi\epsilon} \tan^{-1}\left( \frac{S_w}{4\chi^2\sqrt{1+\frac{S_w}{2\chi^2}}}\right). 
\label{disp_x}
\end{eqnarray}

Then, calling $\xi^2 = \frac{2\chi^2}{S_w}$, \eref{disp_x} becomes

\begin{equation}
\Phi(\xi) = \frac{q}{\pi\epsilon} \tan^{-1} \left( \frac{1}{2\sqrt{\xi^2(1+\xi^2)}}\right),
\label{disp_inter}
\end{equation}

such that the condition $S_w \gg \chi^2$ becomes equivalent to $\xi \ll 1$. So \eref{disp_inter} becomes simply:

\begin{equation}
\Phi(\xi)_{\xi^2 \ll 1} =  \frac{q}{\pi\epsilon} \tan^{-1} \left( \frac{1}{2\sqrt{\xi^2}}\right). 
\end{equation}

Remembering that $\tan^{-1}(\alpha \xi)+\tan^{-1}(\frac{1}{\alpha \xi}) = \frac{\pi}{2}$ for $\xi>0$ then one has:

\begin{equation}
\Phi(\xi) =  \frac{q}{\pi\epsilon} \left[\frac{\pi}{2} - \tan^{-1}\left( 2 \xi \right)\right].
\label{disp_chi}
\end{equation}

In \eref{disp_chi} the term $\tan^{-1}(2\xi)$ can be expanded obtaining: 

\begin{equation}
\Phi(\xi) = \frac{q}{\pi\epsilon} \left[ \frac{\pi}{2} - 2 \xi + \frac{(2\xi)^3}{3} - ...\right].
\label{disp_taylor}
\end{equation}

This last expression, \eref{disp_taylor}, can be truncated at first order of $\xi$ for our large surface. Thus recalling the original variables one arrives at:

\begin{equation}
\Phi(X) = \frac{q}{\pi\epsilon} \left[\frac{\pi}{2} - 2\sqrt{\frac{2}{S_w}} (x_w-X) \right] \propto X.
\label{flux_S}
\end{equation}

\eref{flux_S} is an important results, it demonstrates that the flux of the electric field generated by a particle in a very large surface is proportional to the position of the electron.\\
Now it can be discussed the general problem considered here, i.e. derive a microscopic analysis of the measurement of the total electrical current in a large metallic surface. In order to do that one has to ``enlarge'' the system considering also all the electrons composing the metallic surface, as done described in the main text.\\
Without assuming nothing about the dynamics of the electrons in the metal, one can say that they contribute to the flux of the total electric field as described by \eref{fflux} by superposition principle. One obtains, suppressing the dependence on $x_w$ and making reference to the position of the electron in the device as $X_1$, the following expression:

\begin{equation}
\label{signal_noise}
\Phi(X_1,\mathbf{X}_2,...,\mathbf{X}_N) = \alpha X_1 + \alpha \sum_{k=2}^{N} \Phi(\mathbf{X}_k),
\end{equation}

where the actual Bohmian positions of the particles $\mathbf{X}_i$ have been used and $\alpha$ is a suitable constant.
In \eref{signal_noise} one can clearly see that the total electric flux is due to a contribution from the electron in the system $\propto X_1$ and another due to all the other electrons in the metal.\\
So far, it has been considered that the electron in the active region of the device is not crossing the surface and therefore one gets that the total electric current is due only to the displacement current contribution. So the total current becomes: 

\begin{eqnarray}
I_{T_{S_w}} &\propto& \frac{d \Phi}{dt} =  \frac{d}{dt} \left( \alpha X_1(t) + \alpha \sum_{k=2}^{N}  \Phi(\mathbf{X}_{k}(t))  \right) \nonumber \\
&\propto&  v_{x_1}  +   \sum_{k=2}^{N}  \nabla \Phi(\mathbf{X}_k(t)) \cdot \mathbf{v}_{k}, 
\label{total_current}
\end{eqnarray}

where $v_{x_i} \forall i$ is the $x$-component of the Bohmian velocity of the $i$-particle. One can reasonably assume that, for a large number of particles in the cable, we get $\sum_{k=2}^{N} \langle  \nabla \Phi(\mathbf{X}_k(t)) \cdot \mathbf{v}_{k} \rangle \approx 0$. So, finally we arrive to the result in Eq. (9), 

\begin{equation}
\langle I_{T} \rangle_{S_w} \propto  \langle p_{x_1} \rangle.
\label{total_main_text}
\end{equation}

\eref{total_main_text} shows that the mean value of the total electrical current in a large metallic surface is proportional to the mean value of the momentum ($x$-component, i.e. the component perpendicular to the surface) of the quantum particle in the device. Let us emphasize that we arrive to the same result, Eq. (9) in the letter,  from completely different arguments, without using the Ramo-Shockley-Pellegrini expressions, nor the quantum expression of the particle current density. 

\section{C -Frequency dependence of the weak measurement of the momentum}

Here it is addressed the question on how the frequency on which the measurement is performed influenced the weakness of the measurement. In general a weak measurement requires that: the measurement does not perturb (too much) the wave function, that the information extracted from a single experiment is not reliable and that the mean value corresponds exactly to the mean value of a strong measurement of the same quantity. In Fig. 2a in the letter it has been shown that if one wants precise information about the measured system, the perturbation of the wave function is increased, and vice versa. It has been also proven that the mean value obtain from the simulations with ammeter is equal to the mean value without ammeter (of the system alone), as clearly shown in Fig. 2.

In the letter, we show numerically and explicitly that the weak measurement of the total current can be written in the language of Gaussian measurement Kraus operator as:

\begin{eqnarray}
\hat{I}_{w}(p_w) = C_w \int dp e^{-\frac{(p-p_w)^2}{2\sigma_w}} |p\rangle \langle p|,
\label{weak-def-I} 
\end{eqnarray}

where $p$ is the momentum (x-component) of the particle in the device, which is exactly the form used in the main text.\\

\begin{figure}[h!!!]
\centering
\includegraphics[width=0.84\columnwidth]{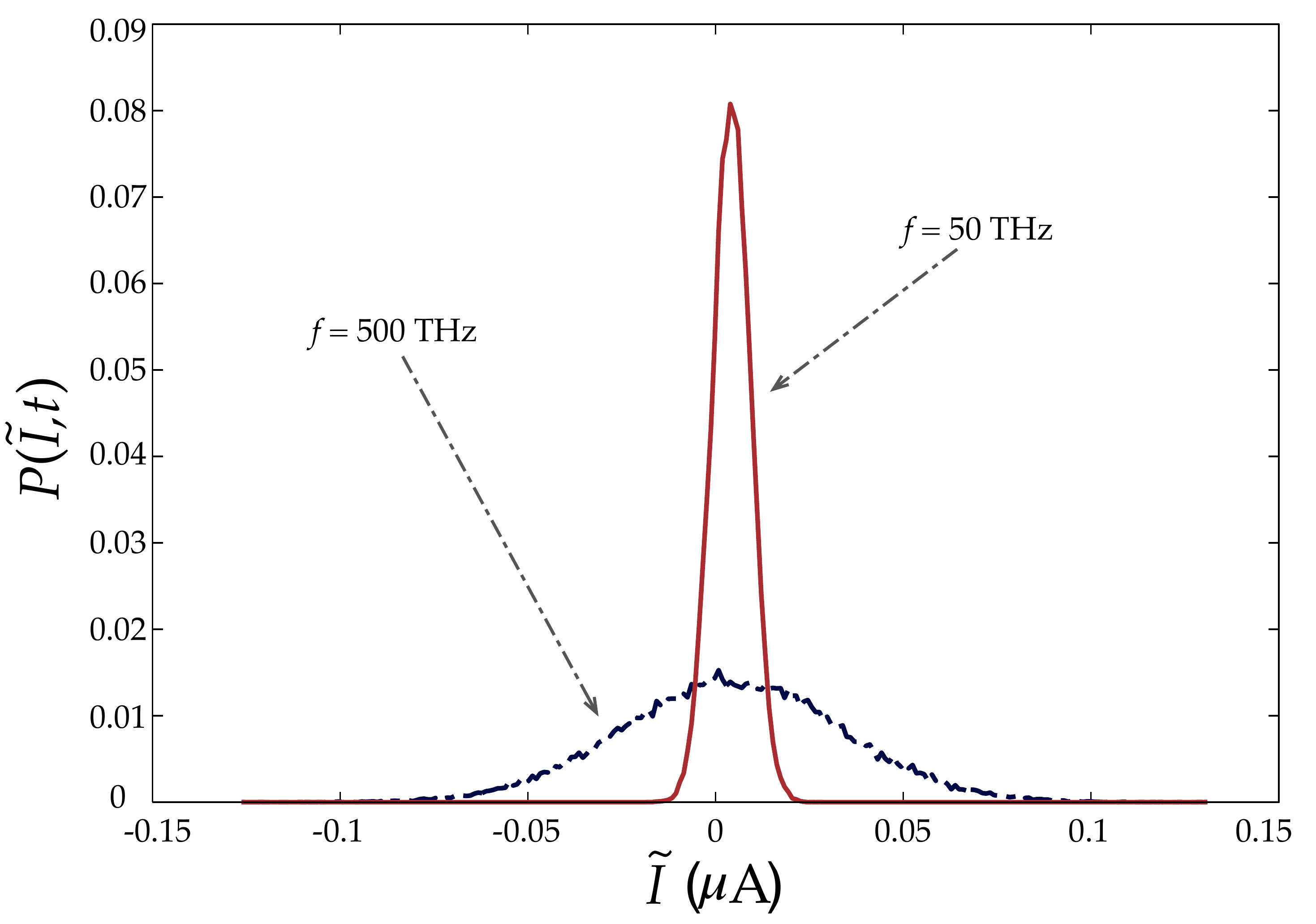}
\caption{Blue dashed line probability distribution of the measured total current at a frequency of $f=500 \cdot 10^{12}\;\;Hz$. Red solid line probability distribution of the measured total current at $f=50 \cdot 10^{12}\;\;Hz$.}
\label{fig-sigmas}
\end{figure}

Now, it will be discussed how the width $\sigma_w$ of the Gaussian measurement Kraus operator changes with the frequency of the current measurement. In \fref{fig-sigmas} it is reported how $\sigma_w$ varies with the frequency of the measurement. It can be seen that lowering the frequency yields to a more precise (mean) information about the system measured. As said in Ref. \cite{footnote1} our ammeter have to work at a frequency $f<1/\tau$ where $\tau$ is the dwell time of the electron in the specific device, otherwise we will not get information about the interference pattern of the wave function in the numerical experiment considered in the main text.

\section{D - Total Current and position measurements on a small Surface}
\label{app3}

In the letter, we have demonstrated through the use of the quantum version of the Ramo-Shockley-Pellegrini theorem that, in the case of a surface $S_s$, where the electrical flux is collected is very small, compared to the distance of the electron from the surface, the measurement of the displacement current can be interpreted as a position measurement. Here we provide an alternative demonstration of Eq. (12) in the letter through the use of the \emph{Conditional} wave function algorithm described in the letter.

We consider the case in which the surface $S_s$, where the electrical flux is collected is very small compared to the distance of the electron from the surface $\chi = x_s-X$. Specifically one has to consider \eref{disp_x} in the limit in which $S_s \ll \chi$ assuming as before that the electron is on the left of the surface, i.e. $\chi>0$. Making the following change of variable $\xi = \frac{S_s}{2\chi^2}$ the previous condition becomes equivalent to study the condition $\xi \ll 1$ for the function

\begin{equation}
\Phi(\xi) = \frac{q}{\pi \epsilon} \tan^{-1}\left( \frac{\xi}{2\sqrt{1+\xi}} \right),
\end{equation}

it can be easily seen that the first order expansion of the last expression is: 

\begin{equation}
\Phi(X) = \frac{q}{4 \pi \epsilon} \frac{S_s}{(x_s -X)^2} \:\:\:\: \text{for} \:\:\:\: X<x_s. 
\end{equation}

The physical interpretation of this term is quite natural: the contribution of a particle to the total flux measured in a small surface is only relevant when the particle is ``near'' the surface. This means that the contribution is relevant when the electron cross the surface. i.e:
\begin{equation}
X \approx x_s+\sqrt{S_s}\;\;\;\;with \;\;\; S_s \rightarrow 0.
\end{equation}

We note that the function \eqref{newdisp} has a discontinuity in the point $X = x_{s}$ of the first kind. This is just an artificial result due to the consideration of the displacement current alone. If we add the particle current in the discussion of  the total flux of the electric field, then the total current has no discontinuity. The particle current is just a delta function centered at $x_s$, i.e. $q \delta(X-x_s)$. 

Thus, since the flux of the displacement current can be interpreted (roughly) as a type of Heaviside step function, its time derivative will be also proportional to a delta function. Therefore,  the total (quantum ensemble) current measured on the small surface $S_s$ will be :

\begin{eqnarray}
\langle I_{T} \rangle_{S_s} &\propto& \langle \delta(x_1-x_s) \rangle = \nonumber \\ 
&=& \int dx_1 \psi^*(x_1,t) \delta(x_1-x_s) \psi(x_1,t),
\label{strong_current}
\end{eqnarray}  

where $\psi(x_1,t)$ is the wave function of the electron in the device. \eref{strong_current} explains the obvious relation that measuring the total current in a small surface provides information whether or not the particle is passed through the surface.  In principle a treatment including the rest of the electrons of the system can be provided but in this case the result obtained does not change; in fact the electron in the device contributes to the total current only when pass through the small surface and then when the interaction with all the others electrons in the metal is strong (the distance between the electron in the quantum system and the metal electrons is very small, implying a strong perturbation ).

Thus it is possible to write the measurement of the total current in a small surface $S_s$ in the language of the Gaussian measurement Kraus operator as:

\begin{eqnarray}
\hat{I}_{s}(x_s) = C_s \int dx e^{-\frac{(x-x_s)^2}{2\sigma_s}} |x\rangle \langle x|,
\label{current-strong}
\end{eqnarray}

which is exactly the form used in the Eq. (12) in the letter. 

\section{E - Classical treatment of the Electric Field}
\label{app4}

The treatment of the electromagnetic interaction between electrons is treated classically in the sense that no quantization of photons is considered. Although it is a typical assumption in most of the high frequency simulations of electron devices, hereafter we provide a detailed justification.  Following Ref. \cite{Electric_Field} we have that an electric field can be treated classically if the following condition is satisfied:

\begin{equation}
| E |^{2} \gg \frac{\hbar}{c^3 \Delta t^4 \epsilon_0}.
\label{cond_ef}
\end{equation}

This condition arise from counting the number of photons in a time interval $\Delta t$. In our simulations we have that the time interval is the inverse of the frequency reported in Fig. 2, so $\Delta t = 2 \cdot 10^{-14}\;s $. For this value of the interval of time $\Delta t$ the right hand side of \eref{cond_ef} is:

\begin{equation}
\frac{\hbar}{c^3 \Delta t^4 \epsilon_0} \approx 2,7 \cdot 10^{6}\;\; \frac{N^2}{C^2}.
\end{equation}

We do not have, from the simulations, the direct information about the modulus of the electric field, but it can be estimated from the flux of the electric field through the surface used for detecting the displacement current. In particular from the simulations reported in the main text we have:

\begin{equation}
|\bar{E}| = \frac{\Phi}{S_w},
\end{equation}

where $|\bar{E}|$ is the modulus of the mean electric field over the entire surface. In the simulations reported in the main text we have used a surface of $S_w = 10^{-13}\;m^2$. The flux of the electric field, through the same surface, is $\Phi \approx 7 \cdot 10^{-10}\; V\cdot m$. Thus for these values of the parameter we have

\begin{eqnarray}
|\bar{E}|^2 \approx  5 \cdot 10^{7}\;\; \frac{N^2}{C^2}, 
\end{eqnarray}

satisfying the relation reported in \eref{cond_ef} and thus justifying the use of a classical electrical field in our simulations. We remind that, due to the dimensions of our device, we can use the (non-local) Coulomb potential as a reliable approximation in our simulations. This is a standard approximation for transport in mesoscopic systems.


\begin{thebibliography}{99}
\bibitem{footnote1}
We emphasize that the ammeter, as any other electronic devices, works as a low-pass filter. Therefore, the current measured by the ammeter, after the electronic filtering, will be $\tilde{I}_{freq}(t) = 1/T \int_{t}^{t+T} \tilde{I}(t') dt'$, i.e. the current measured at frequency $f=1/T$. Our discussion is valid for frequencies $f > 1/\tau$, where $\tau = L_x/v$ is the dwell time of the electron in the device, which depends on the length of the device $L_x$ and on the velocity $v$ of the electron. In the text we have defined $\tilde{I} \equiv \tilde{I}_{freq}(t)$ to simplify the notation.
\bibitem{footnote7} The spatial dependence of the vector potential can be neglected in front of the spatial dependence of the scalar potential for active regions of nanometric dimensions until frequencies of few hundreds of THz.
\bibitem{footnote3} 
In particular, we will concentrate only in the interaction between the electron of the quantum system and $N_P$ electrons in the probe which are close enough to the system (the coulomb force is inversely proportional to the squared distance).
\end{thebibliography}
\end{document}